\documentclass[twocolumn,showpacs,aps,floatfix,superscriptaddress]{revtex4}
\pdfoutput=1 
\usepackage{amsmath,amssymb,graphicx,eucal,bm}
\usepackage{multirow}

\begin{document}

\title{Crystal Growth Inside an Octant}

\author{Jason Olejarz}
\affiliation{Center for Polymer Studies, and Department of Physics, Boston University, Boston, MA 02215, USA}
\author{P.~L.~Krapivsky}
\affiliation{Department of Physics, Boston University, Boston, MA 02215, USA}

\begin{abstract}
We study crystal growth inside an infinite octant on a cubic lattice.  The growth proceeds through the deposition of elementary cubes into inner corners. After re-scaling by the characteristic size, the interface becomes progressively more deterministic in the long-time limit.  Utilizing known results for the crystal growth inside a two-dimensional corner, we propose a hyperbolic partial differential equation for the evolution of the limiting shape.  This equation is interpreted as a Hamilton-Jacobi equation which helps in finding an analytical solution. Simulations of the growth process are in excellent agreement with analytical predictions.  We then study the evolution of the sub-leading correction to the volume of the crystal, the asymptotic growth of the variance of the volume of the crystal, and the total number of inner and outer corners. We also show how to generalize the results to arbitrary spatial dimension.
\end{abstract}

\pacs{68.35.Fx, 05.40.-a, 02.50.Cw }

\maketitle

\section{Introduction}

Shapes of growing objects have fascinated humans from the dawn of time. The understanding of these shapes is crucial for numerous technological applications. Microscopic processes underlying growth phenomena are usually stochastic, with rules depending on the detailed local structure of the interface, so it is not surprising that even the simplest growth rules lead to interfaces which are seemingly impossible to describe theoretically. An important relatively recent theoretical insight is the realization that microscopic details often play a secondary role. This has led to the devising of {\em continuum} descriptions for fluctuations of growing interfaces. The most well-known such framework was initiated by Kardar, Parisi, and Zhang (KPZ) who proposed a continuum theory based on a nonlinear stochastic partial differential equation \cite{KPZ}, arguably the simplest equation accounting for the crucial growth ingredients --- nonlinearity, stochasticity, irreversibility, and locality. The KPZ equation provides a unifying framework for probing fluctuations in a large class of growing interfaces.  A comprehensive description of fluctuations of one-dimensional growing interfaces has subsequently emerged (see \cite{HZ95,KK10} and references therein),  and a key recent breakthrough is a solution of the 1+1 dimensional KPZ equation \cite{KPZ_rev}.  Crystal growth typically occurs in three dimensions, however. Microscopic growth models and continuum theories are straightforward to formulate in arbitrary dimension, yet in 2+1 dimensions the KPZ equation remains inaccessible to current analytical approaches. Therefore fluctuations of two-dimensional growing interfaces are still poorly understood. 

Fluctuations characterize the local structure of the interfaces, but they tell nothing about the overall shape of an interface, more precisely the shape on the scale greatly exceeding length scales associated with fluctuations. Such overall shapes, known as limiting shapes, usually become clearly identifiable in the long-time limit. In the simplest case when growth begins from a flat substrate, the limiting shape is trivial --- the interface remains on average flat, and only speed and fluctuations matter.  In most applications the interfaces are curved. Curved limiting shapes have been analytically determined only in a few cases. For instance, the limiting shape is still unknown for the two-dimensional Eden-Richardson growth model, although it has been proved that the limiting shape exists and that it is roughly but not exactly circular \cite{eden}; in contrast, fluctuations of the interface of Eden clusters are understood (and belong to the 1+1 dimensional KPZ universality class).

There are almost no analytical results for the limiting shapes of curved two-dimensional growing interfaces. All known tractable examples correspond to {\em anisotropic} growth in 2+1 dimensions. The term anisotropic means that the first two dimensions, the transversal directions along the interface, arise in a greatly distinct manner. An anisotropic 2+1 dimensional growth model can usually be reformulated as a collection of identical solvable 1+1 dimensional growth models with some non-intersecting condition between neighboring interfaces. One solvable anisotropic growth model is the 2+1 dimensional Gates-Westcott model \cite{GW95} which mimics vicinal growing surfaces; this model has been solved by a free-fermion mapping~\cite{Prahofer}. Average interface profiles are also known for two other anisotropic 2+1 dimensional growth models~\cite{RajeshDhar,Borodin}. In the most basic isotropic growth models limiting shapes are not known. 

One of the first non-trivial limiting shapes was found in the context of the corner growth model~\cite{rost}. In this model, one starts with an infinite empty corner, namely with a quadrant on the square lattice; the growth occurs by deposition of $1\times 1$ squares into available inner corners. The limiting shape consists of two ballistically receding half-lines constituting the original boundary which are connected by a piece of parabola
\begin{equation} 
\label{2d_shape}
\begin{cases}
      x=0  & y>t\\
      \sqrt{x} +  \sqrt{y}  =  \sqrt{t} & 0<x,y<t\\
      y=0   &x >t
\end{cases}      
\end{equation}
This two-dimensional corner growth model has also played a crucial role in the analytical description of the statistics associated with 1+1 dimensional KPZ growth~\cite{BDJ,J00}. 

In this paper we consider the natural generalization of the corner growth process to three dimensions, where the growth occurs inside an octant, and to higher dimensions. We also study the behavior of integral characteristics, e.g. the volume of the crystal (we look at the average and the standard deviation), and the total numbers of inner and outer corners on the interface. 

The rest of this paper is organized as follows. The growth process is defined in Sect.~\ref{mod} where we also recall known two-dimensional results.  In Sect.~\ref{3D} we investigate the three-dimensional case, namely we consider crystal growth on the cubic grid inside the octant. We propose an evolution equation describing the asymptotic evolution of the interface. In Sect.~\ref{analysis} we interpret this equation as the Hamilton-Jacobi equation, and we use an equivalent description based on the canonical Hamilton equations to derive the solution. Generalizations to higher dimensions are discussed in Sect.~\ref{higher}. The behavior of the average volume of the crystal (both the leading asymptotic and the sub-leading correction) and volume fluctuations are investigated in Sect.~\ref{volume}. The growth of the total numbers of inner and outer corners is analyzed in Sect.~\ref{corners}. In Sect.~\ref{concl} we summarize what we understand,  emphasize the remaining challenges, and discuss open problems.

\section{The Model}
\label{mod}

\begin{figure}[ht]
\begin{center}
\includegraphics[width=0.375\textwidth]{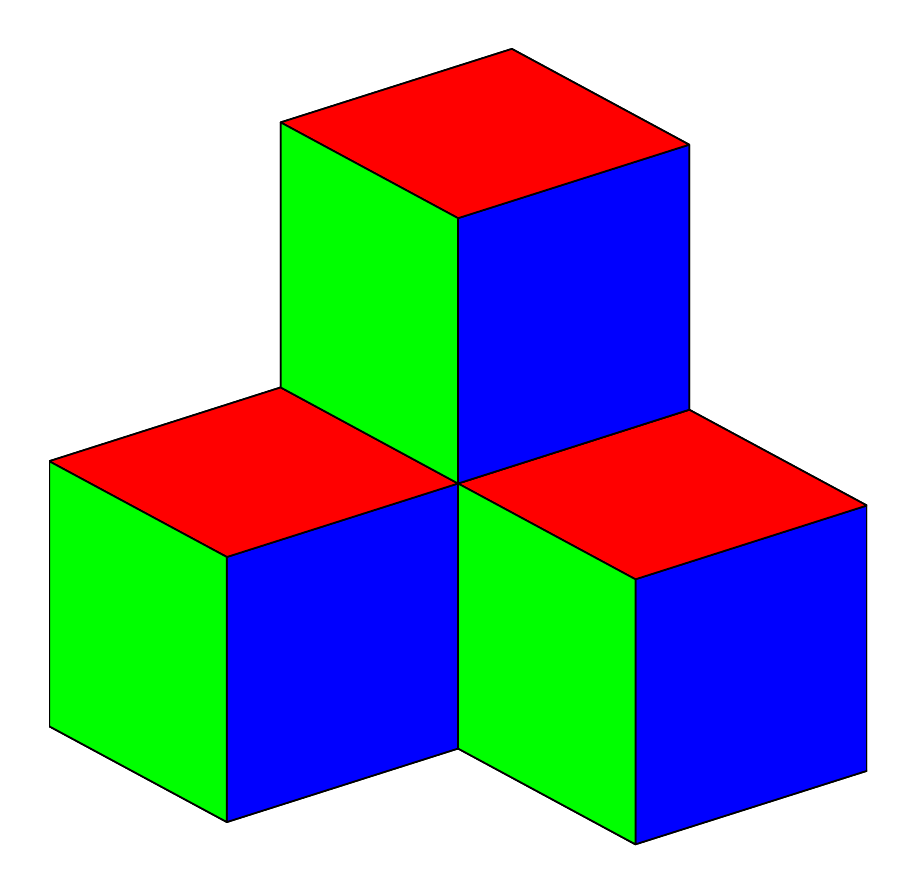}
\caption{\small A three-dimensional crystal of volume 4. A new cube can be deposited to 6 places. }
\label{Fig:illust_3d}
  \end{center}
\end{figure}

Consider an infinite corner, viz. a positive octant in the cubic lattice, which is initially empty. The growth process begins at time $t=0$ and proceeds by depositing elementary $1\times 1\times 1$ cubes into available inner corners. We set the deposition rate to unity without loss of generality. Initially there is one inner corner available, so the smallest non-empty crystal is unique. After this first deposition event, there are three available inner corners that can accommodate the next cube, so the uniqueness is lost. For the crystal shown in Fig.~\ref{Fig:illust_3d} there are 6 places to insert a new cube. 

The above dynamics can be defined in arbitrary spatial dimension. The models are of course lattice models; more specifically they are defined on hyper-cubic lattices. (One can consider other lattices, but then one should choose a different infinite initially empty region depending on lattice structure.) 

\begin{figure}[ht]
\begin{center}
\includegraphics[width=0.375\textwidth]{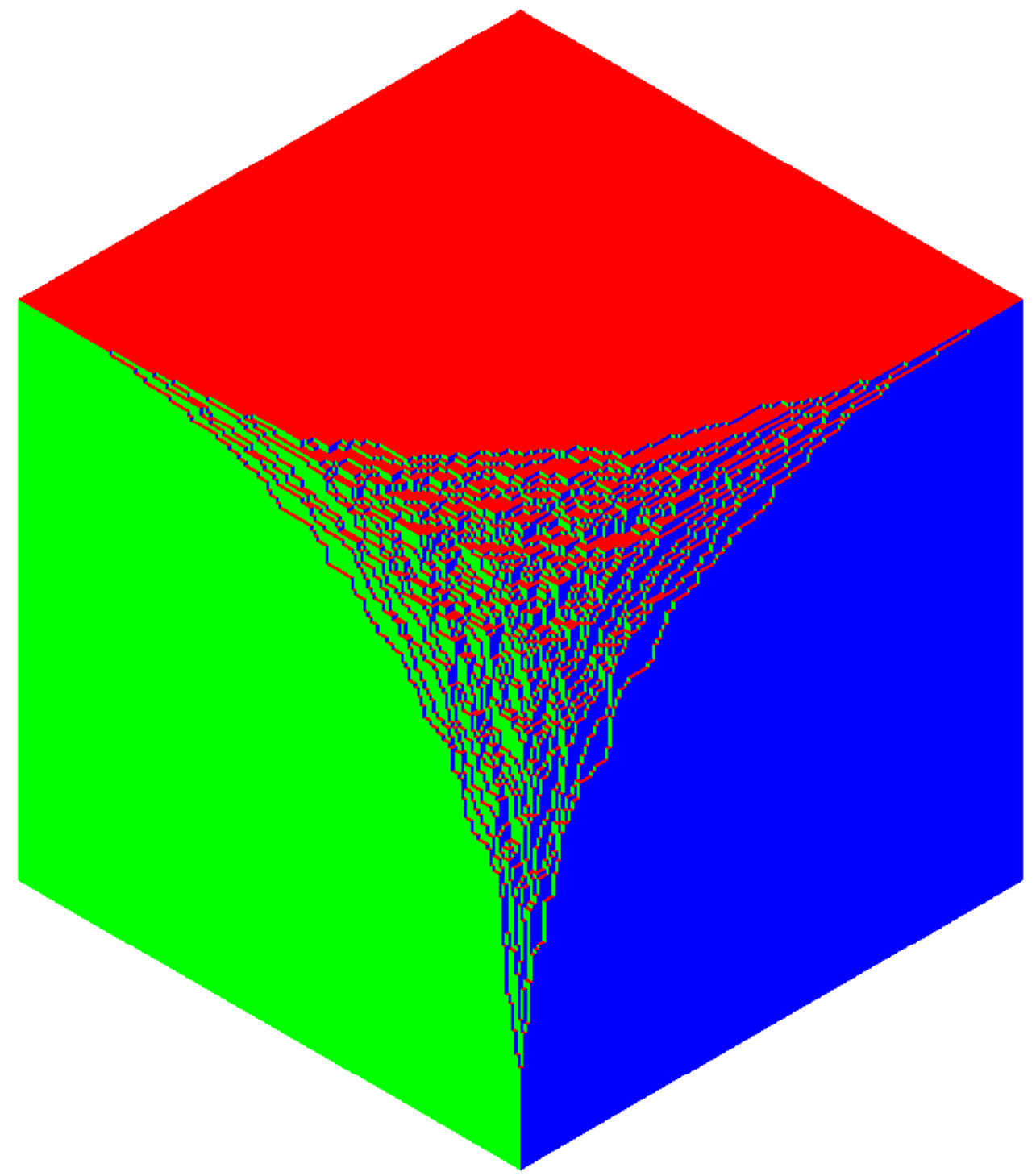}
\caption{\small Melting of a three-dimensional crystal initially occupying a negative octant; a view from the $(1,1,1)$ direction.}
\label{Fig:crystal_3d}
  \end{center}
\end{figure}

The interpretation is a matter of taste: Rather than talking about crystal growth through the deposition of cubes into inner corners, we can think about crystal melting through the desorption of cubes from outer corners of an initially fully filled octant. These two processes are dual. A typical realization of the melting process is presented in Fig.~\ref{Fig:crystal_3d}. The interface is stochastic, yet as the crystal grows (or equivalently as the melted volume increases), the interface becomes smoother and ultimately approaches a deterministic limiting shape. 

\begin{figure}
\centering
\includegraphics[scale=0.36]{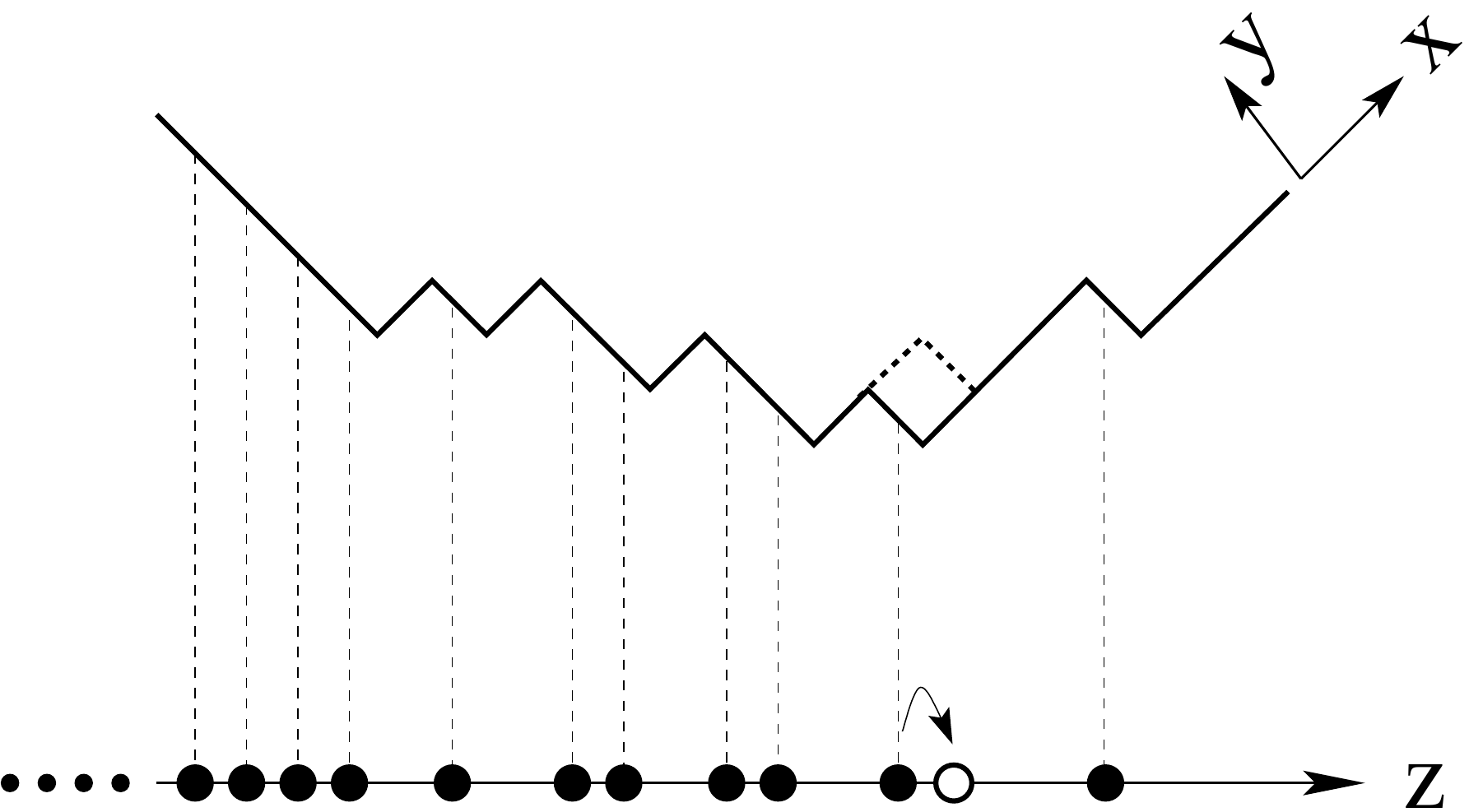}
\caption{An example of the interface of the two-dimensional crystal rotated by $\pi/4$ and the corresponding particle configuration. A deposition event is shown, and the corresponding hop of a particle to a neighboring vacant site  is highlighted.}
\label{csp}
\end{figure}

The two-dimensional corner growth process can be understood by a mapping onto the one-dimensional totally asymmetric simple exclusion process (TASEP), that is a collection of particles undergoing a biased random walk under the constraint that there is at most one particle per site~\cite{rost}. (This mapping is also helpful in computing  fluctuations of the interface~\cite{BDJ,J00}.)  The representation in terms of the TASEP becomes evident after one rotates the corner counter-clockwise by an angle of $\pi/4$ around the origin and then projects the interface onto the one-dimensional lattice (see Fig.~\ref{csp}) in such a way that bonds of the interface are identified with sites on the one-dimensional lattice. We now put a particle on a site (leave a site empty) if the corresponding segment on the interface goes along the co-diagonal (diagonal). Each site on the one-dimensional lattice is occupied by at most one particle, the particles hop to the right with unit rate, and the lattice gas is clearly identical to the TASEP. The average density $\rho(z,t)$ evolves according to the (inviscid) Burgers equation
\begin{equation*}
\frac{\partial \rho}{\partial t} +\frac{\partial [\rho(1-\rho)]}{\partial z} = 0\,.
\end{equation*}
Solving this Burgers equation and expressing the limiting shape through the density  \cite{book} one arrives at a remarkably simple parabolic limiting shape \eqref{2d_shape}.

In three dimensions, the corner growth model can be mapped onto an infinite set of coupled TASEPs in the plane, also known as the `zigzag model' \cite{Tamm}.  Unfortunately, no exact solutions are known for
such planar interacting particle processes. The generalization to higher dimensions also seems exceedingly difficult. We therefore choose a different strategy; namely, we try to directly write an evolution equation for the limiting shape without using intermediate mappings onto a particle process. 

In two dimensions, this program can be fulfilled, viz. the limiting shape $y(x;t)$ satisfies the evolution equation~\cite{Liggett,spohn,karma}
\begin{equation}
\label{eqn:yt-2d}
y_t=\frac{y_x}{y_x-1}
\end{equation}
where $y_t=\frac{\partial y}{\partial t}, ~y_x=\frac{\partial y}{\partial x}$. The solution to Eq.~\eqref{eqn:yt-2d} is indeed given by \eqref{2d_shape}. Our task is to generalize Eq.~\eqref{eqn:yt-2d} to three and higher dimensions. 

\section{Three Dimensions}
\label{3D}

We haven't derived a generalization Eq.~\eqref{eqn:yt-2d} to three dimensions. In such a situation, one can proceed in a less systematic manner by guessing an evolution equation. The criteria are the simplicity and  the validity of the basic symmetry properties. More precisely, we seek an evolution equation for $z(x,y;t)$ that
\begin{enumerate}
\item Reduces to the proper analogs of Eq.~\eqref{eqn:yt-2d} on the boundaries $x=0$ and $y=0$, viz. $z_t=z_y/(z_y - 1)$ on the first boundary and $z_t=z_x/(z_x - 1)$ on the second.
\item Is invariant in form under the $x\leftrightarrow z$ and $y\leftrightarrow z$ coordinate exchanges.
\end{enumerate}

Needless to say, we anticipate that the governing equation is a first-order partial differential equation (PDE) of the form $z_t=F(z_x,z_y)$ similar to the two-dimensional case.  The simplest guess for the right-hand side is the product of terms similar to those appearing in Eq.~\eqref{eqn:yt-2d}: 
\begin{equation}
\label{wrong}
   z_t =  \frac{z_x}{z_x - 1} \, \frac{z_y}{z_y - 1} \,. 
\end{equation}
On the boundaries $x=0$ and $y=0$ we have $z_x=-\infty$ and $z_y=-\infty$, and hence \eqref{wrong} reduces to $z_t=z_y/(z_y - 1)$ and $z_t=z_x/(z_x - 1)$ on these boundaries. Yet Eq.~\eqref{wrong} does not possess the required invariance under the exchange of variables. 

We eventually found an equation satisfying the above criteria. This equation has a very unusual form 
\begin{equation}
\label{unique}
   z_t =  \frac{z_x}{z_x - 1} \, \frac{z_y}{z_y - 1} \left[1- \frac{1}{z_x + z_y}\right].
\end{equation}
Our search was facilitated by numerical experiments in which we studied the movement of various planes. (More precisely, we initially investigated a generalized version of the hypercube stacking model of Forrest and Tang \cite{FT} which allows for tuning the slopes $z_x$ and $z_y$ of the flat interface.) Any plane moves (on average) with a constant speed. We thus measure the speed and compare it with the prediction of Eq.~\eqref{unique}. For the plane $z_x=z_y=-1$ (this is the plane orthogonal to the (1,1,1) direction) we found $z_t\approx 0.378$ which is an excellent agreement with the prediction $z_t=\frac{3}{8}$ implied by \eqref{unique}. We looked at several other planes, e.g. for the plane $z_x= - \frac{1}{2},\; z_y = -2$ the measured velocity is again within $1\%$ of the prediction $z_t=\frac{14}{45}$ implied by  \eqref{unique}. 

\subsection{Arguments in favor of \eqref{unique}}

Consider a one-parameter family of evolution equations
\begin{equation}
\label{family}
   z_t =  \frac{z_x}{z_x - 1} \, \frac{z_y}{z_y - 1}  \, \frac{z_x+z_y+\lambda}{z_x + z_y}  \,. 
\end{equation}
This family contains Eqs.~\eqref{wrong} and \eqref{unique} as special cases. The invariance under the exchange of $x$ and $y$ is manifest. Further, equation \eqref{family} reduces to $z_t=z_y/(z_y-1)$ [respectively to $z_t=z_x/(z_x-1)$]  when $x=0$ [respectively $y=0$]. Hence we only need to test the invariance under the exchange of $x$ and $z$. We use standard relations between the derivatives 
\begin{equation}
\label{zx_1}
  z_t = - \frac{x_t}{x_z}\,,\quad z_x =  \frac{1}{x_z}\,,\quad z_y = - \frac{x_y}{x_z} \,. 
\end{equation}
By inserting \eqref{zx_1} into \eqref{family} we arrive at
\begin{equation*}
   x_t =  \frac{x_z}{x_z - 1} \, \frac{x_y}{x_y - 1}  \, \frac{x_y-\lambda x_z- 1}{x_y + x_z}  \,. 
\end{equation*}
On the other hand, the invariance under the exchange of $x$ and $z$ requires that we should obtain
\begin{equation*}
   x_t =  \frac{x_z}{x_z - 1} \, \frac{x_y}{x_y - 1}  \, \frac{x_y+x_z + \lambda}{x_y + x_z}  \,. 
\end{equation*}
Comparing these two equations we conclude that the only member of the family of equations \eqref{family}  that satisfies the invariance requirements corresponds to $\lambda=-1$, which is precisely the evolution equation \eqref{unique}.  In particular, Eq.~\eqref{wrong}, which corresponds to $\lambda=0$, is {\em not} invariant under the exchange of $x$ and $z$.

Instead of a one-parameter family of evolutionary equations \eqref{family} let us now analyze an infinite-parameter family of evolutionary equations
\begin{equation}
\label{inf_family}
   z_t =  \frac{z_x}{z_x - 1} \, \frac{z_y}{z_y - 1} 
   \left[1+\sum_{n\geq 1}\frac{\lambda_n}{(z_x + z_y)^n}\right]. 
\end{equation}
The invariance under $x\leftrightarrow z$ exchange leads to 
\begin{eqnarray*}
&&1+\frac{\lambda_1}{x_y + x_z}+\frac{\lambda_2}{(x_y + x_z)^2}+\frac{\lambda_3}{(x_y + x_z)^3}
+\ldots\\
&=&1-\frac{1}{x_y + x_z}-\frac{(\lambda_1+1)x_z}{x_y + x_z}
+ \frac{\lambda_2 x_z^2}{(x_y-1)(x_y + x_z)}\\
&-&\frac{\lambda_3  x_z^3}{(x_y-1)^2(x_y + x_z)}+\ldots
\end{eqnarray*}
This is valid for arbitrary $x_y,\, x_z$ only when $\lambda_1=-1$ and $\lambda_n=0$ for $n\geq 2$. 

We can further generalize the class of equations \eqref{inf_family}, namely by replacing the term in the square brackets in \eqref{inf_family} with an arbitrary Laurent series
\begin{equation}
\label{inf_family_inf}
   z_t =  \frac{z_x}{z_x - 1} \, \frac{z_y}{z_y - 1} 
   \sum_{n=-\infty}^\infty\frac{\lambda_n}{(z_x + z_y)^n}  \,. 
\end{equation}
Terms with $n<0$ cannot be present, however, since on the boundaries $x=0$ and $y=0$ we must recover $z_t=z_y/(z_y-1)$ and $z_t=z_x/(z_x-1)$; this requirement additionally fixes the parameter $\lambda_0=1$ and hence we are back to \eqref{inf_family}. Therefore Eq.~\eqref{unique} is the only appropriate evolution equation among the family of equations \eqref{inf_family_inf}.

\subsection{Caveats}

The evolution equation \eqref{unique} is not a unique evolution equation satisfying the necessary requirements. For instance, we have found another equation
\begin{equation}
\label{unique_bad}
   z_t =  \frac{z_x}{z_x - 1} \, \frac{z_y}{z_y - 1} \left[1 +  \frac{1}{z_xz_y -z_x - z_y}\right]
\end{equation}
that obeys all necessary conditions. Equation \eqref{unique_bad} looks more complicated than \eqref{unique}, although it can be re-written in a very simple (even if a bit unusual) form if we replace derivatives by their reciprocal values:
\begin{equation}
\label{grow_3}
\frac{1}{z_t} = 1 - \frac{1}{z_x} - \frac{1}{z_y} \,. 
\end{equation}
This equation admits a very simple analytical solution which is a straightforward generalization of the solution in the two-dimensional setting:
\begin{equation}
\label{3_parabola}
\sqrt{x} + \sqrt{y} + \sqrt{z} = \sqrt{t} \,. 
\end{equation}
This is verified by a direct substitution. The analytical solution of  \eqref{unique} is different from \eqref{3_parabola} as we shall see. 

A comparison of predictions of \eqref{unique} and \eqref{unique_bad} is strongly in favor of Eq.~\eqref{unique}. For instance, consider the volume of the crystal. For the surface \eqref{3_parabola} corresponding to the evolution equation \eqref{unique_bad}, the volume is $t^3/90$. It is much more difficult to compute the volume corresponding to the evolution equation \eqref{unique}. An exact solution to Eq.~\eqref{unique} is rather cumbersome; namely it is parametric, so one must compute an unwieldy integral. The answer is  
\begin{equation}
\label{volume_3d}
V_3=v_3 t^3\,,\qquad v_3 = \frac{3\pi^2}{2^{11}}=0.014457428321908...
\end{equation}
The amplitude substantially exceeds $\frac{1}{90}$ corresponding to the interface \eqref{3_parabola}; numerically $v_3\approx 0.01472(3)$. 

Further, let us look at the intersection of the interface and the diagonal [in the $(1,1,1)$ direction]. One can numerically determine this quantity with a good precision. Analytically, this point corresponds to 
\begin{equation}
\label{diagonal}
x=y=z=w t
\end{equation}

For the surface \eqref{3_parabola}, we have $w=\frac{1}{9}$, while $w=\frac{1}{8}$, on the interface predicted by Eq.~\eqref{unique}. Interestingly, we can extract $w=\frac{1}{8}$ without knowing the limiting shape. The high symmetry of the diagonal implies that  
\begin{equation}
\label{diagonal_diff}
x_y=x_z=y_z=y_x=z_x=z_y=-1 
\end{equation}
on the diagonal \eqref{diagonal}. Indeed, all the derivatives in \eqref{diagonal_diff} must be equal due to symmetry. To establish the numerical value we use $z_x=x_z$ together with the identity $x_z=1/z_x$ to conclude that $(z_x)^2=1$ from which (the derivatives are obviously negative) we arrive at \eqref{diagonal_diff}. 

Plugging \eqref{diagonal_diff} into \eqref{unique} we find that $z_t=\frac{3}{8}$ on the diagonal. Projecting the vector $(0,0,z_t)$ onto the diagonal $(1,1,1)$ direction we find that the distance of the diagonal point on the interface from the origin is equal to $\frac{3}{8}\frac{1}{\sqrt{3}}t$. To determine $x=y=z$ we need to project again; this yields $w=\frac{3}{8}\frac{1}{\sqrt{3}}\frac{1}{\sqrt{3}}=\frac{1}{8}$. 

Numerically $w\approx 0.1261(2)$. This result is close to the theoretical prediction $w=0.125$ and clearly differs from $w=\frac{1}{9}=0.111\ldots$ that characterizes the interface \eqref{3_parabola}. 

In principle, one can use equations \eqref{unique} and \eqref{unique_bad} as building blocks to obtain one-parameter families of equations satisfying the necessary requirements. Two such families are obtained by an additive and a multiplicative combination of \eqref{unique} and \eqref{unique_bad}. An additive family is
\begin{equation}
\label{alt}
z_t =\frac{z_x}{z_x\!-\!1}\, \frac{z_y}{z_y\!-\!1} \left[1\!-\!\frac{1+c}{z_x+z_y} \!-\! \frac{c}{z_xz_y \!-\!z_x \!-\! z_y}\right]\!,
\end{equation}
and a multiplicative family is
\begin{equation}
\label{multiplicative}
z_t =
\left[\frac{1-\frac{1}{z_x+z_y}}{\big(1-\frac{1}{z_x}\big) \big(1-\frac{1}{z_y}\big)}\right]^{1+c}
\left[1 - \frac{1}{z_x} - \frac{1}{z_y}\right]^c\!.
\end{equation}
For the additive class of evolution equations~\eqref{alt}, the choice $c\approx 0.079$ provides the best fit for the numerically determined value of $w$ \cite{vel}; for the multiplicative class of evolution equations~\eqref{multiplicative}, the optimal choice of the mixing parameter is $c\approx 0.074$. The corner interface growth is presumably described by a simple equation that does not contain an anomalously small mixing parameter.  This  in conjunction with our numerical results suggest that Eq.~\eqref{unique} describes corner interface evolution; at the very least, the true evolution equation is not an ugly deformation like \eqref{alt} or \eqref{multiplicative} with a very small mixing parameter. 

Using equations \eqref{unique} and \eqref{grow_3} separately, one can construct a few more families of invariant equations. One such one-parameter series family 
\begin{equation*}
\label{zt_n}
z_t =\frac{z_x}{z_x-1}\, \frac{z_y}{z_y-1} \left[1-\frac{1}{z_x+z_y}\right]
\frac{(1-z_x-z_y)^n}{1+(-z_x)^n+(-z_y)^n}
\end{equation*}
extends the presumably correct equation \eqref{unique} corresponding to $n=1$ to arbitrary real $n$. Setting $n=1+\log_3(8w)$ would match the observed value of $w$; for $w=0.126$, one gets  $n\approx 1.00725$. A similar extension of \eqref{grow_3} is
\begin{equation*}
\frac{1}{z_t} = \left[1 - \frac{1}{z_x} - \frac{1}{z_y}\right]\frac{1+(-z_x)^n+(-z_y)^n}{(1-z_x-z_y)^n}
\end{equation*}
Choosing $n=3+\log_3 w$ would match the observed value of $w$, so for $w=0.126$ one gets  $n\approx 1.114$. 

We can even construct multi-parameter families of evolution equations by simply multiplying any number of additional factors of the form
\begin{equation*}
\frac{(1-z_x-z_y)^{n_i}+\alpha_i(z_xz_y)^{n_i/3}}{1+(-z_x)^{n_i}+(-z_y)^{n_i}+\beta_i(z_xz_y)^{n_i/3}}
\end{equation*}
onto the right-hand side of~\eqref{alt} or~\eqref{multiplicative}, where each $n_i$, $\alpha_i$, and $\beta_i$ is a free parameter.  Each such factor strictly preserves the invariance properties of the evolution equation.  In principle, we could obtain nearly perfect theoretical agreement over the entire simulated interface profile by suitably tuning parameters in these equations.  The trade-off is that the growth equations are becoming quite unsightly.  That our numerical simulations persistently show such minute discrepancies from a beautiful analytical description~\eqref{unique} is puzzling \cite{OKRM2012}.  Even for careful simulations of the hypercube stacking model \cite{comment}, tiny (yet apparently significant) inconsistencies between our simple equation~\eqref{unique} and simulation data persist.  Theoretically explaining the precise source of these discrepancies is an intriguing open question.

\subsection{Simulation results}

For $2+1$ dimensional corner growth, we saved the simulated interface profile at times $t_i=20000/2^{(9-i)/2}$ with $i=0,\ldots,9$, giving a total of ten data points.  We ran forty independent realizations of the growth process until at least time $t=7071$.  We then continued running twenty of those realizations until at least time $t=10000$, and we ran ten of those realizations all the way to $t=20000$.  Each of our measurements of a quantity at time $t_i$ is therefore an ensemble average.  The error bar for a measurement is computed as the standard error of the measured quantity over all forty, twenty, or ten realizations that were run for at least that long.

It is possible that discrepancies between our simulation results and the predictions that follow from Eq.~\eqref{unique}  can be attributed to rather slow convergence to the asymptotic state.  Flat and curved interface geometries have been proven to have differing fluctuation statistics in $1+1$ dimensions \cite{PS2000}, and we are operating in the much-less-understood $2+1$ dimensional setting.  
A similarly slow convergence to asymptotic behavior occurs in various well-understood one-dimensional growth models (see e.g. Refs.~\cite{farnudi,ferrari,ferreira_1d}).  For example, for 1+1 dimensional corner growth, the intersection of the interface with the $(1,1)$ direction evolves according to~\cite{BDJ,J00,KK10}
\begin{equation}
\label{fluct_2}
x(t) = \frac{t}{4} + t^{1/3}\, \xi\,,
\end{equation}
where $\xi$ is a stationary random variable with $\langle \xi\rangle >0$.
Thus averaging over many realizations gives an effective velocity
$w_\text{eff}-\frac{1}{4} \sim t^{-2/3}$.

For growth inside a three-dimensional corner, we therefore anticipate that $w_\text{eff} - \frac{1}{8}\sim t^{-\alpha}$, with a certain (theoretically unknown) exponent $\alpha$.  Very extensive simulations for flat interfaces in 2+1 dimensions
indicate that $\alpha$ is close to 0.77~\cite{KrugMeakin,reis,odor,ferreira}.  On the other hand, extrapolation from our simulations for $t\alt 20000$ suggests that $\alpha\approx 0.74$.  This difference in exponent estimates suggests that $t=20000$ is still outside the long-time regime for growth inside a three-dimensional corner.  It has proved difficult to obtain consistent estimates of the KPZ scaling exponents from simulations even in the simpler case of growing flat interfaces in $2+1$ dimensions (see e.g. \cite{Wolf,Chate,HH12}).  This slow approach to the asymptotic behavior can be the source of the discrepancy between our simulation results and the theoretical prediction \eqref{unique} for the interface profile.

\begin{figure}[ht!]
\begin{center}
  \includegraphics*[width=0.375\textwidth]{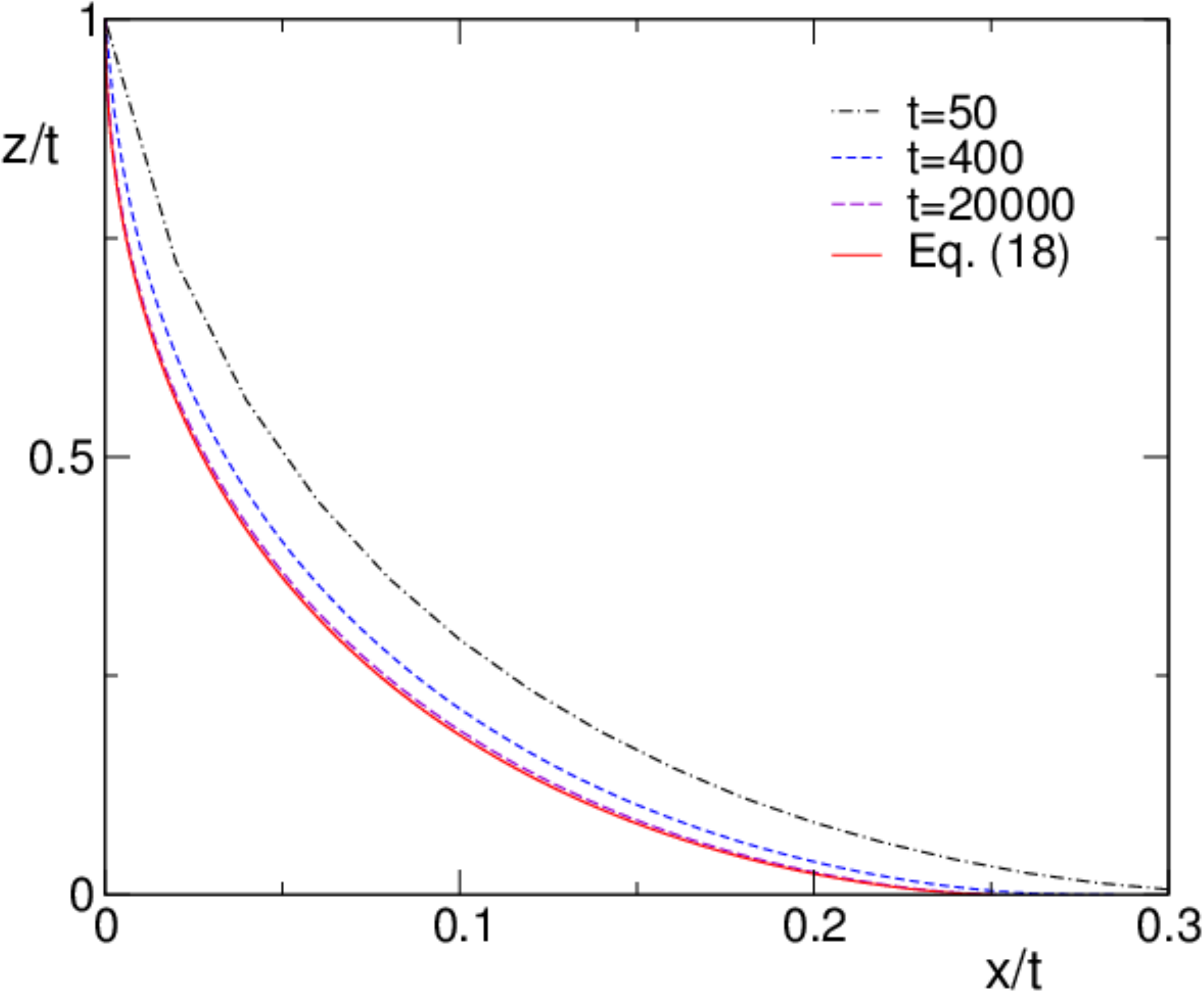}
  \caption{\small Scaled interface profile, $z/t$  versus $x/t$, along the diagonal $x=y$ at different times.}
\label{fig:diag}
  \end{center}
\end{figure}

\begin{figure}[ht!]
\begin{center}
  \includegraphics*[width=0.375\textwidth]{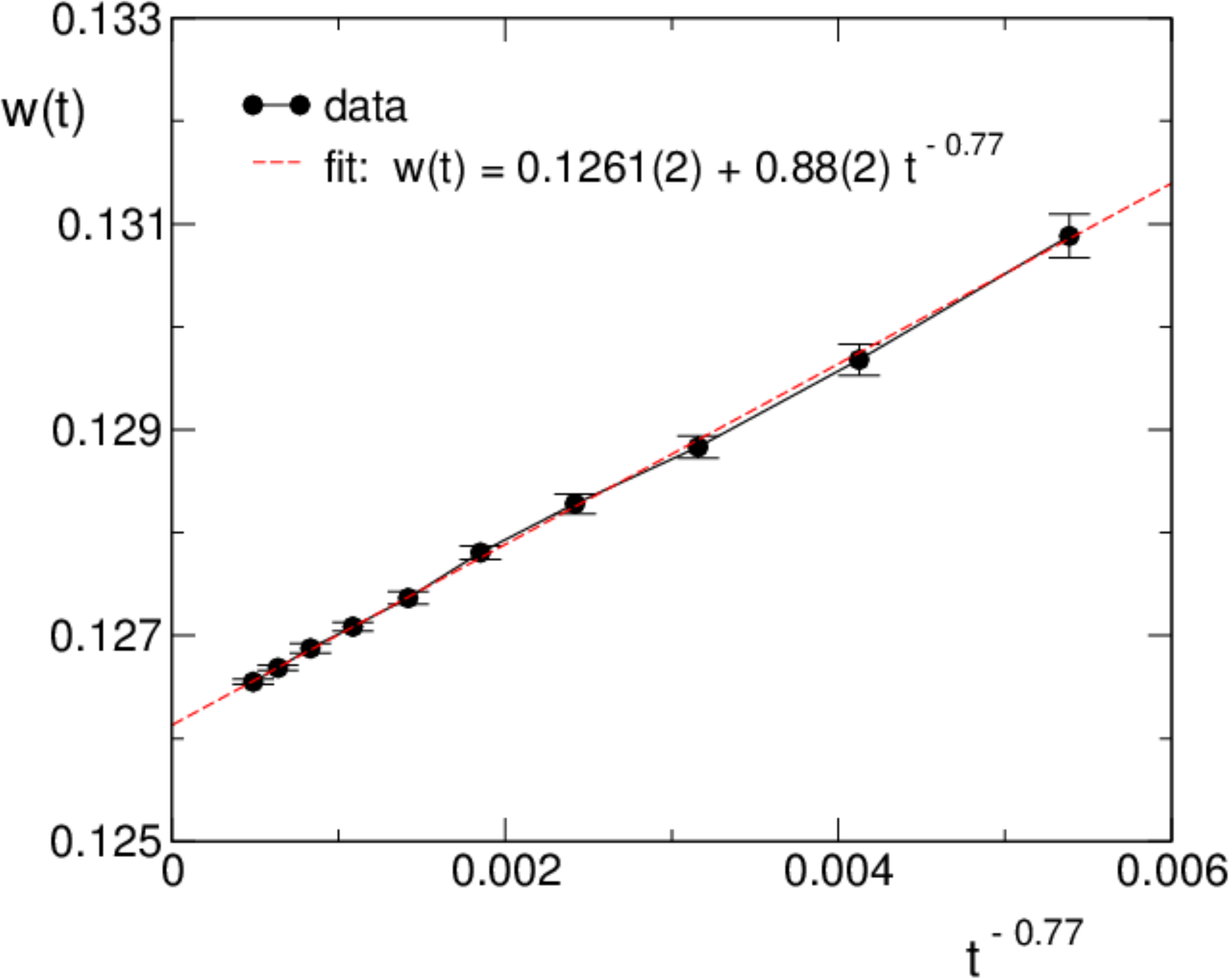}
  \caption{\small Convergence of the diagonal interface speed versus $t^{-0.77}$.  This quantity appears to 
  settle at a value slightly greater than $0.126$, which is roughly $0.9\%$ off from our prediction $0.125$.}
\label{fig:convergence}
  \end{center}
\end{figure}

As an additional numerical test, consider the intersection of the interface with the plane $x=y$. From our analytical solution of \eqref{unique}, we obtain 
\begin{equation}
\label{shape:diag}
\frac{x}{t}=\frac{1}{2}\frac{z}{t}-\frac{3}{4}\left(\frac{z}{t}\right)^{2/3}+\frac{1}{4}
\end{equation}
which agrees well with simulations (Fig.~\ref{fig:diag}).

\section{Analytical Determination of the Limiting Shape}
\label{analysis}

In this section we solve the evolution equation \eqref{unique} and determine the limiting shape predicted by Eq.~\eqref{unique}. Equation \eqref{unique} is solvable since it is a first-order (hyperbolic) PDE. Such equations can be treated using the method of characteristics \cite{Char}. The method of characteristics is especially efficient in applications to linear and quasi-linear hyperbolic PDEs. Equation \eqref{unique} is fully non-linear and in such cases the analysis involving the method of characteristics tends to be cumbersome. Fortunately, there is a shortcut in the present situation: We can employ the Hamilton-Jacobi formalism \cite{Gelfand,Arnold}. A trick is to interpret $z=z(x,y;t)$ as an action. Then \eqref{unique} becomes the Hamilton-Jacobi equation, $z_t+H=0$, with Hamiltonian 
\begin{equation}
\label{H:pp}
H = -\frac{p_1}{p_1-1}\frac{p_2}{p_2-1}\left[1-\frac{1}{p_1+p_2}\right].
\end{equation}
Here $p_1$ and $p_2$ are momenta, i.e., the spatial derivatives of the action 
\begin{equation}
\label{pp}
p_1 = \frac{\partial z}{\partial x}\equiv z_x, \qquad p_2 = \frac{\partial z}{\partial y}\equiv z_y \,. 
\end{equation}
The canonical Hamilton equations for coordinates are
\begin{equation}
\label{xy}
\frac{dx}{dt} = \frac{\partial H}{\partial p_1}\,, \qquad 
\frac{dy}{dt} = \frac{\partial H}{\partial p_2} \,. 
\end{equation}
The canonical Hamilton equations for momenta show that both momenta are constant (this is obvious since the Hamiltonian \eqref{H:pp} does not depend on the coordinates). Plugging \eqref{H:pp} into \eqref{xy} we arrive at
\begin{eqnarray}
\label{xt}
\frac{dx}{dt} = A &\equiv& \frac{1}{(p_1-1)^2}\frac{p_2}{p_2-1}\left[1-\frac{1}{p_1+p_2}\right]\nonumber\\
&&-\frac{p_1}{p_1-1}\frac{p_2}{p_2-1}\frac{1}{(p_1+p_2)^2}
\end{eqnarray}
and 
\begin{eqnarray}
\label{yt}
\frac{dy}{dt} =  B &\equiv& \frac{1}{(p_2-1)^2}\frac{p_1}{p_1-1}\left[1-\frac{1}{p_1+p_2}\right]\nonumber\\
&&-\frac{p_1}{p_1-1}\frac{p_2}{p_2-1}\frac{1}{(p_1+p_2)^2} \,. 
\end{eqnarray}

In our concrete problem, the action variable $z$ plays the same role as $x$ and $y$; the separate treatment of $x,\, y,$ and $z$ in Eq.~\eqref{unique} is just a matter of choice. To determine $z$ we integrate the Hamilton-Jacobi equation $z_t+H=0$ to yield $z=-Ht+F(x,y)$, and then recalling \eqref{pp} we fix $F(x,y)=p_1x+p_2y$. Integrating Eqs.~\eqref{xt}--\eqref{yt} and combing these results with $z=-Ht+p_1x+p_2y$ we get
\begin{equation}
\label{ABC_xyz}
\frac{x}{t} = A, \quad
\frac{y}{t} = B, \quad
\frac{z}{t} = C
\end{equation}
with $A(p_1,p_2)$ and $B(p_1,p_2)$ defined in \eqref{xt} and \eqref{yt}, and $C(p_1,p_2)$ given by
\begin{equation}
\label{zt}
C = Ap_1+Bp_2-H
\end{equation}

\begin{figure}[ht]
\begin{center}
\includegraphics[width=0.375\textwidth]{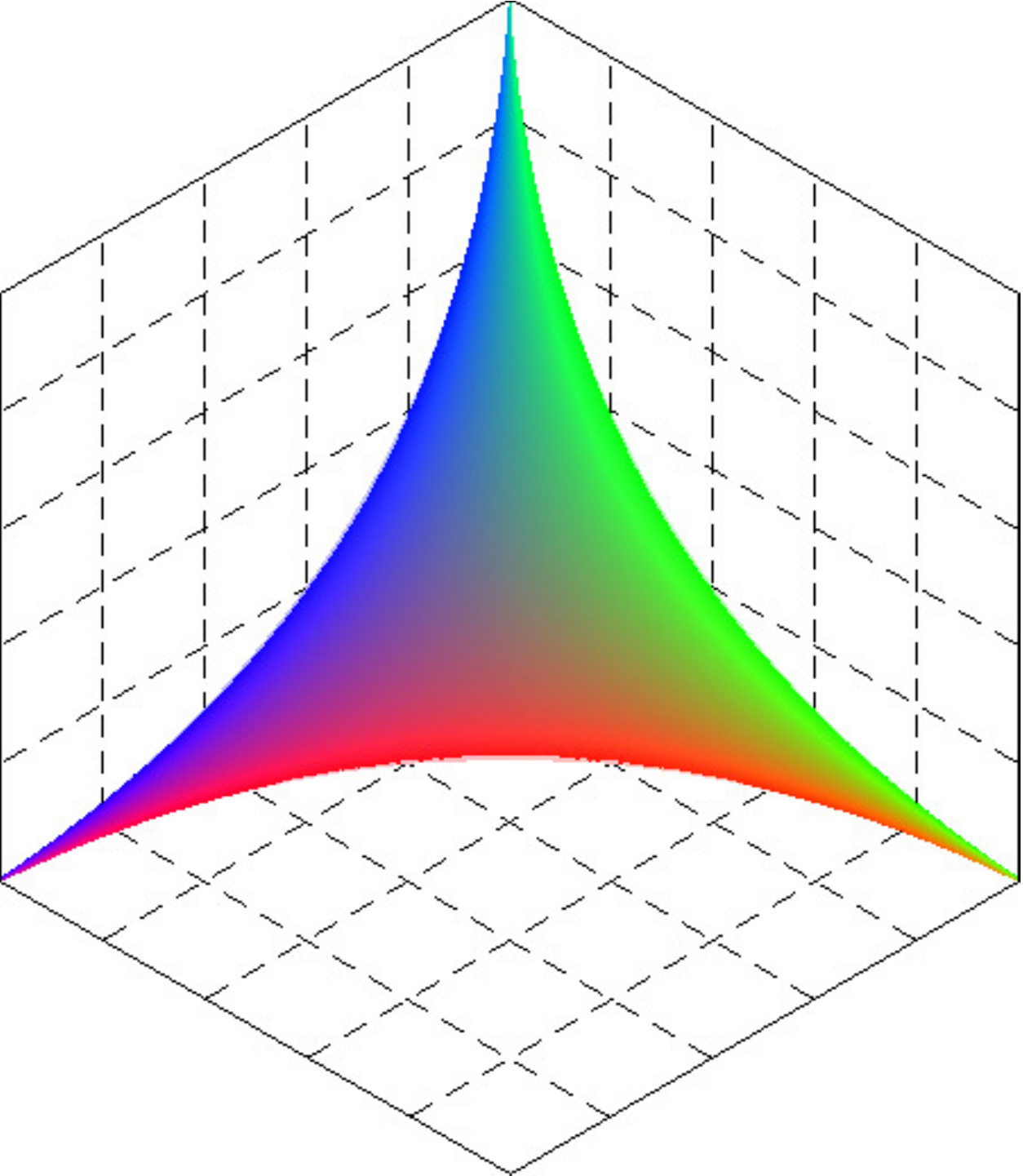}
\caption{\small The interface \eqref{ABC_xyz}.}
\label{Fig:shape_3d}
  \end{center}
\end{figure}

For any fixed time, Eq.~\eqref{ABC_xyz} gives an exact parametric representation of the limiting shape of the interface; the plot of the interface is presented in Fig.~\ref{Fig:shape_3d}. More precisely, \eqref{ABC_xyz} represents the non-trivial part of the interface, with parameters varying in the range $-\infty<p_1,p_2<0$. For every fixed $(p_1,p_2)$ we can think of $x(t)$, $y(t)$, and $z(t)$ as a point growing along a line, this line is merely a characteristic of the original PDE. 

As a check of consistency we note that for $p_2=-\infty$, we have $A=(p_1-1)^{-2}, ~B=0, ~C=p_1^2(p_1-1)^{-2}$. In other words, the non-trivial part of the intersection of the interface \eqref{ABC_xyz} and the $y=0$ plane is 
\begin{equation*}
\frac{x}{t} = \frac{1}{(p_1-1)^2}\,, \qquad \frac{z}{t} = \frac{p_1^2}{(p_1-1)^2}
\end{equation*}
from which we get $\sqrt{x}+\sqrt{z}=\sqrt{t}$, i.e. we recover the two-dimensional result [cf. \eqref{2d_shape}]. 

It seems impossible to exclude the parameters $(p_1,p_2)$ and find an explicit expression for the interface \eqref{ABC_xyz} in terms of $x/t, y/t, z/t$. Some lines on the surface defined by \eqref{ABC_xyz} can be explicitly presented, however. Consider for instance the intersection of the surface \eqref{ABC_xyz} and the plane $x=y$. On this plane $p_1=p_2$ and therefore this line has a parametric representation
\begin{equation}
\label{xxz:Q}
\frac{x}{t} = \frac{y}{t} = \frac{1}{2}\,P^3+\frac{3}{4}\,P^2\,, \quad
\frac{z}{t} = (P+1)^3 
\end{equation}
where $P=(p_1-1)^{-1}$.  Excluding the parameter $P$ we recast \eqref{xxz:Q} into the announced result \eqref{shape:diag}. 

To compute the volume of the crystal
\begin{equation}
\label{V3}
V_3 = \int\int dx\,dy\,z
\end{equation}
we need to integrate over the region 
\begin{equation*}
0<x<t,\quad 0<y<t, \quad \sqrt{x}+\sqrt{y}\leq \sqrt{t} \,. 
\end{equation*}
Using the exact solution \eqref{ABC_xyz} we reduce \eqref{V3} to $V_3=v_3 t^3$, with amplitude $v_3$ given by
\begin{equation*}
v_3 = \int_{-\infty}^0  \int_{-\infty}^0 dp_1\,dp_2\,\, C\,\frac{\partial(A,B)}{\partial(p_1,p_2)} \,. 
\end{equation*}
Using expressions for $A(p_1,p_2),  B(p_1,p_2),  C(p_1,p_2)$ from Eqs.~\eqref{xt}, \eqref{yt}, \eqref{zt} and computing the Jacobian $\frac{\partial(A,B)}{\partial(p_1,p_2)}$ we reduce $v_3$ to a cumbersome but elementary integral. We computed this integral with the help of \emph{Mathematica} and obtained the value given in  \eqref{volume_3d}.

\section{Higher Dimensions}
\label{higher} 

We want to generalize Eqs.~\eqref{eqn:yt-2d} and \eqref{unique}. In four dimensions the governing equation for the height $h(x,y,z;t)$ has the form 
\begin{equation}
\label{grow_4d}
h_t=\frac{\big(1-\frac{1}{h_x+h_y}\big)\big(1-\frac{1}{h_y+h_z}\big)\big(1-\frac{1}{h_z+h_x}\big)}
{\big(1-\frac{1}{h_x}\big)\big(1-\frac{1}{h_y}\big)\big(1-\frac{1}{h_z}\big)
\big(1-\frac{1}{h_x+h_y+h_z}\big)}
\end{equation}
This equation is manifestly symmetric in $x$, $y$, $z$ and it reduces to proper equations when one of the derivatives ($h_x,\, h_y$, or $h_z$) goes to $-\infty$. Therefore it suffices to test that \eqref{grow_4d} is invariant under the exchange of $x$ and $h$. Substituting 
\begin{equation}
\label{hx_1}
h_t = -\frac{x_t}{x_h}\,,~ h_x = \frac{1}{x_h}\,,~ h_y = -\frac{x_y}{x_h}\,,~ h_z = -\frac{x_z}{x_h}
\end{equation}
we indeed establish the required invariance. 

Generally in $d+1$ dimensions, the evolution equation for the height $h(x_1,\ldots,x_d; t)$ reads 
\begin{equation}
\label{grow_d}
h_t=\prod_{a=1}^d
\prod_{1\leq i_1<\ldots<i_a\leq d}\left(1-\frac{1}{h_{i_1}+\ldots+h_{i_a}}\right)^{(-1)^a}
\end{equation}
where $h_i\equiv \frac{\partial h}{\partial x_i}$. 

Let us first determine the diagonal point 
\begin{equation*}
x_1=\ldots=x_{d}=h=w_{d} t
\end{equation*}
The derivatives in \eqref{grow_d} are $h_1=\ldots=h_{d}=-1$ at this point and therefore 
\begin{equation*}
h_t = \prod_{a=1}^{d} \left(1+\frac{1}{a}\right)^{(-1)^a \binom{d}{a}}
\end{equation*}
at the diagonal point. The same argument as before gives 
\begin{equation}
\label{wd}
w_d = \frac{1}{d+1} \prod_{a=1}^{d} \left(1+\frac{1}{a}\right)^{(-1)^a \binom{d}{a}}
\end{equation}
The already known values, $w_1=1/4$ and $w_2=1/8$, are followed by 
\begin{equation*}
\frac{3^4}{4^{5}}\,,~~
\frac{3^{10}}{4^{10}}\,,~~
\frac{3^{18}\, 5^6}{2^{47}}\,,~~\frac{3^{28}\, 5^{21}}{2^{98}}\,,~~\frac{3^{28}\, 5^{56}\,7^8}{2^{202}}\,,~~\frac{5^{126}\,7^{36}}{2^{399}}
\end{equation*}

The interface defined by Eq.~\eqref{grow_d} apparently lies above the hyper-surface
\begin{equation}
\label{d_parabola}
\sqrt{x_1}+\ldots+\sqrt{x_{d+1}} = \sqrt{t}, \quad h\equiv x_{d+1}
\end{equation}
that would be a solution if the governing equation for the interface were given by the analog of \eqref{grow_3}
\begin{equation}
\label{grow_D}
\frac{1}{h_t}=1-\frac{1}{h_1}-\ldots-\frac{1}{h_{d}}
\end{equation}
which also satisfies the required invariance properties. Using \eqref{wd} we found that  $w_d > (d+1)^{-2}$, where the latter value is implied by \eqref{grow_D}, in dimensions $2\leq d\leq 50$. We believe that this bound is generally valid, and it seems that the asymptotic behaviors are very different, namely $w_d\approx \frac{0.1}{d+1}$ provides a good fit of the large $d$ behavior of $w_d$ given by \eqref{wd}, although we haven't deduced this asymptotic.

We ran 400 independent realizations of $4d$ corner growth.  We measured the middle point on the interface at times $t_i=1024/2^{(9-i)/2}$ with $i=0,\ldots,9$, so that the velocity at each of the ten time values was averaged over forty independent runs.  Our data yield $w_3 \approx 0.0806$, within roughly $2\%$ of the prediction $3^4/4^5$ of Eq.~\eqref{wd} and $29\%$ greater than the prediction $1/4^2$ based on \eqref{d_parabola}.  At the least, the interface defined by Eq.~\eqref{grow_d} provides a much more accurate description of $3+1$ dimensional corner growth than does \eqref{d_parabola}.

To determine an exact solution of Eq.~\eqref{grow_d} we use again the Hamilton-Jacobi technique. In four dimensions, for instance, the Hamiltonian is
\begin{equation}
H=-\frac{\big(1-\frac{1}{p_1+p_2}\big)\big(1-\frac{1}{p_2+p_3}\big)\big(1-\frac{1}{p_3+p_1}\big)}
{\big(1-\frac{1}{p_1}\big) \big(1-\frac{1}{p_2}\big)\big(1-\frac{1}{p_3}\big)
\big(1-\frac{1}{p_1+p_2+p_3}\big)}
\end{equation}
The non-trivial part of the interface is
\begin{equation}
\label{xh_AB}
 \frac{x_j}{t} = A_j, \quad j=1,2,3; \qquad \frac{h}{t}     =  B
\end{equation}
Here
\begin{eqnarray}
\label{AAA}
\frac{A_j}{H} &=& \frac{1}{p_j(1-p_j)}+\frac{1}{(p_1+p_2+p_3)(1-p_1-p_2-p_3)}\nonumber\\
&+&\frac{1}{(p_j+p_{j+1})(p_j+p_{j+1}-1)}  \nonumber\\
&+& \frac{1}{(p_j+p_{j-1})(p_j+p_{j-1}-1)}
\end{eqnarray}
with $j=1,2,3$ (the indexes $j\pm 1$ are taken modulo 3). Further, 
\begin{equation}
\label{Bp}
\begin{split}
\frac{B}{H} =&~\frac{1}{1- p_1}+\frac{1}{1- p_2}+\frac{1}{1- p_3}-1\\
&~ +\frac{1}{1-p_1-p_2-p_3}\\
&~ +\frac{1}{p_1+p_2-1} + \frac{1}{p_2+p_3-1} + \frac{1}{p_3+p_1-1}
\end{split}
\end{equation}

\begin{figure}[ht]
\begin{center}
\includegraphics[width=0.375\textwidth]{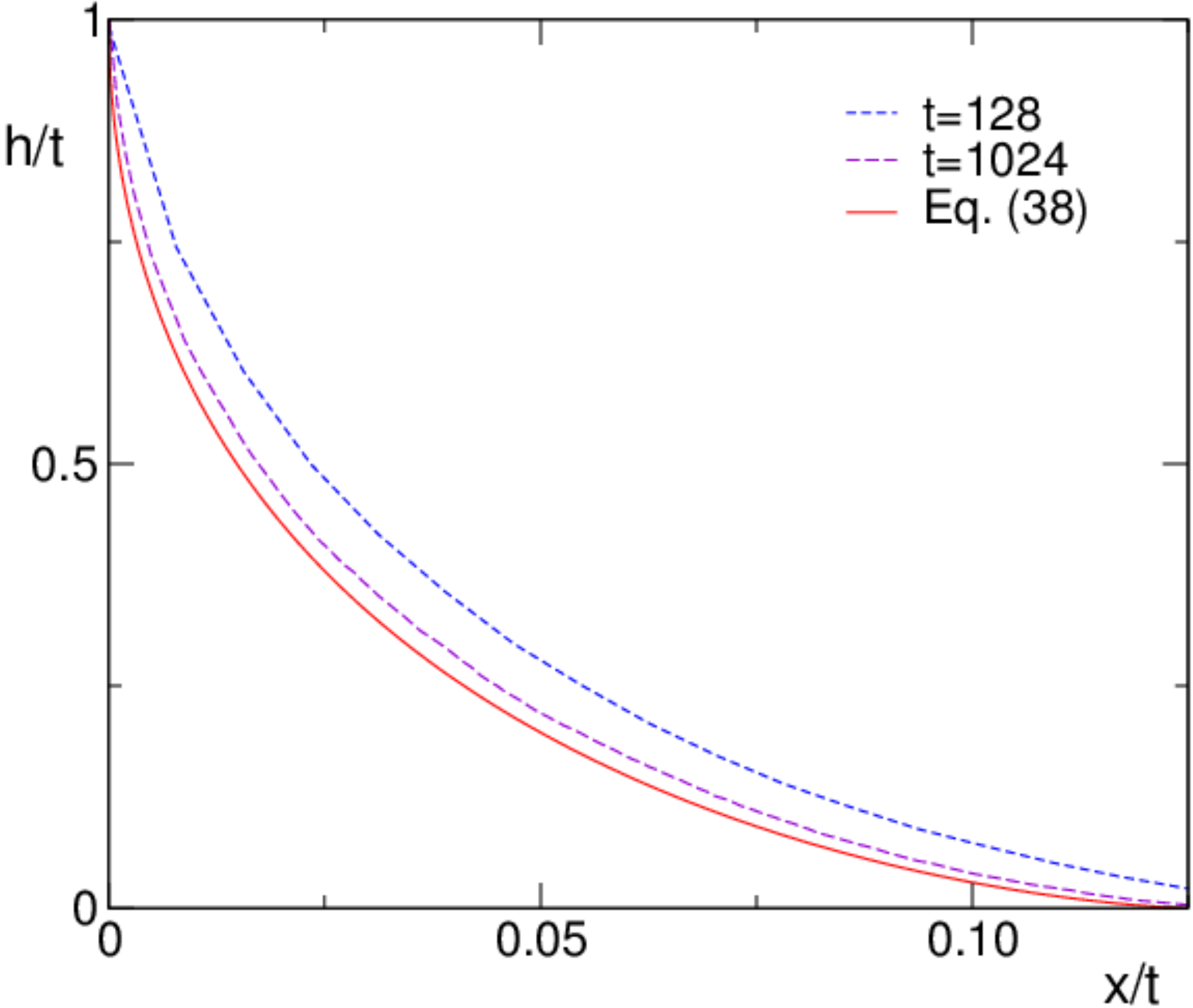}
\caption{\small The height $h/t$ of the $4d$ growing crystal versus $x/t$ with the constraint $x=y=z$; we averaged over 100 realizations for $t=128$ and over  10 realizations for $t=1024$.}
\label{Fig:xxxh}
  \end{center}
\end{figure}

\begin{figure}[ht!]
\begin{center}
  \includegraphics*[width=0.375\textwidth]{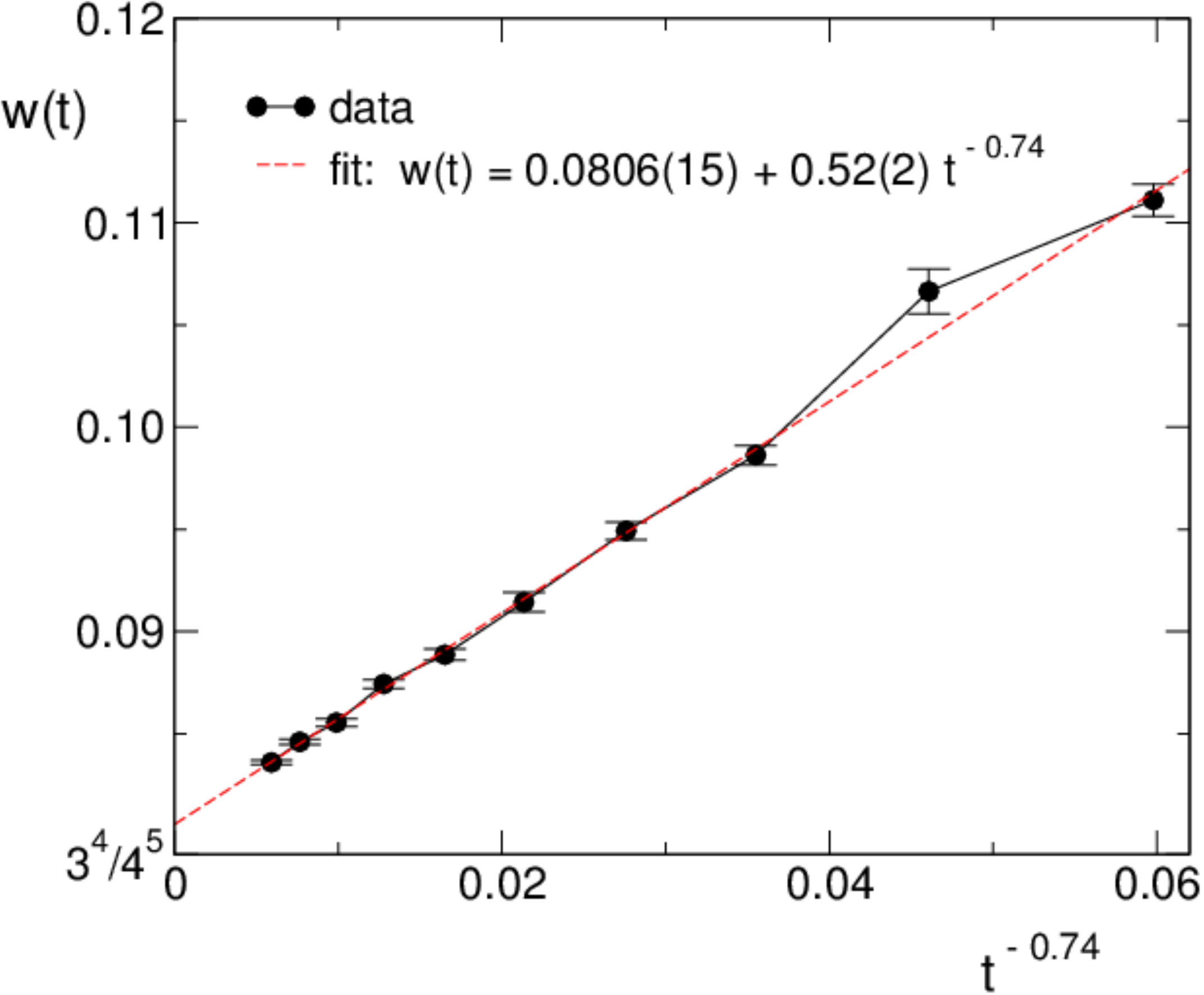}
\caption{\small Convergence of the diagonal interface speed for $4d$ corner growth versus $t^{-0.74}$, where the exponent $-0.74$ gives the best linear fit over the simulated time values.  This quantity appears to approach $0.0806$ as $t\rightarrow\infty$, which is strikingly close to our  prediction $3^4/4^5=0.0791\ldots$.}
\label{fig:4d_convergence}
  \end{center}
\end{figure}

The intersection of this interface with the two-dimensional plane $x_1=x_2=x_3$ is a line that can be parametrically represented as 
\begin{equation}
\label{xxxh_diag}
\begin{split}
& \frac{x}{t} = \frac{(11q^2+6q+1)(2q+1)^2}{8(q+1)^4 (3q+1)^2}\\
&\frac{h}{t} = \frac{9q^4(2q+1)^2}{4(q+1)^4 (3q+1)^2}
\end{split}
\end{equation}
where $q=-p_1=-p_2=-p_3$ varies on the interval $(0,\infty)$. The plot of the line \eqref{xxxh_diag} is given in Fig.~\ref{Fig:xxxh}.

\section{Volume Fluctuations}
\label{volume}

In previous sections, we investigated the limiting shape of the surface of the growing crystal. The growth laws are stochastic, so at any time there are deviations from the deterministic limiting shape. Fluctuations of the interface are rather fully understood only in two dimensions (the growth inside the quadrant), and even in that case fluctuations of the shape were mostly probed (see \cite{KPZ_rev} and references therein). Here we briefly discuss fluctuations of the volume of the crystal in arbitrary dimension. 

The growing crystal has a typical size which scales linearly with time $t$. Therefore the average volume is 
$\langle V\rangle = v_dt^d$ in the leading order. To estimate the sub-leading correction to the average volume and the asymptotic behavior of the variance of the crystal volume, we use heuristic arguments to argue that the average volume grows as
\begin{equation}
\label{Vd_av}
\langle V\rangle = v_dt^d+a_dt^{d-1+\zeta/z}+\ldots
\end{equation}
while the variance grows according to 
\begin{equation}
\label{Vd_var}
\langle V^2\rangle_c\equiv \langle V^2\rangle - \langle V\rangle^2  = b_dt^{d-3+(d+3)/z}
\end{equation}
in the leading order. Here $\zeta$ and $z$ are the well-known exponents \cite{HZ95} which are defined through the growth law 
\begin{equation}
\label{w}
W \sim t^{\zeta/z}
\end{equation}
for the width of the interface and the growth law
\begin{equation}
\label{ell}
\ell\sim t^{1/z}
\end{equation}
for the correlation length. In our crystal growth problem, the interface follows the Kardar-Parisi-Zhang (KPZ) growth laws. The corresponding exponents $\zeta$ and $z$ are related through the KPZ formula \cite{HZ95} 
\begin{equation}
\label{KPZ_exp}
\zeta + z =2
\end{equation}
The exact values are known only in $d=1+1$ dimensions: $\zeta = \tfrac{1}{2}, ~ z = \tfrac{3}{2}$.  Equations \eqref{Vd_av} and \eqref{Vd_var} also contain positive amplitudes $a_d$ and $b_d$ which are analytically unknown even in two dimensions.

To establish \eqref{Vd_av} one makes a natural assumption that the average width scales according to \eqref{w}. The volume is $v_dt^d$ in the leading order, and since the area of the interface scales as $t^{d-1}$, the extra-volume grows as $Wt^{d-1}$ resulting in the sub-leading correction in Eq.~\eqref{Vd_av}. 

To establish \eqref{Vd_var} we divide the interface into patches of size equal to the correlation length. The number of such patches is $(t/\ell)^{d-1}$. Within each patch, the extra-volume is $W\ell^{d-1}$. The deviation of the volume from its average value is therefore the sum $\sum \pm W\ell^{d-1}$ containing $(t/\ell)^{d-1}$ terms. This gives an estimate for the variance $\langle V^2\rangle_c\sim (t/\ell)^{d-1}(W\ell^{d-1})^2$. Using \eqref{w}--\eqref{ell} we obtain  
\begin{equation*}
\langle V^2\rangle_c\sim t^{d-1+(d-1)/z+2\zeta/z}
\end{equation*}
and recalling relation \eqref{KPZ_exp} we arrive at \eqref{Vd_var}. 

In three dimensions, Eq.~\eqref{Vd_av} reduces to
\begin{equation}
\label{V3_av}
\langle V\rangle = v_3t^3+a_3t^{1+2/z}+\ldots
\end{equation}
and \eqref{Vd_var} turns into 
\begin{equation}
\label{V3_var}
\langle V^2\rangle_c  = b_3t^{6/z}
\end{equation}
Careful numerical estimates for the dynamical exponent $z$ characterizing two-dimensional interfaces are given in Ref.~\cite{HH12}.  Numerically, we are dealing with substantial finite-time effects.  Still, our estimates for the lowest-order correction exponent for $\langle V\rangle$ [where we get $~\approx -0.73$ while \eqref{V3_av} gives $t^{-3}\langle V\rangle - v_3\sim t^{-2(1-1/z)}$] and the exponent for $\langle V^2\rangle_c$ [where we get $\approx 4$ while \eqref{V3_var} predicts the exponent $6/z$] are consistent with prior numerical estimates \cite{HH12} of the exponent $z$.

\section{Inner and Outer Corners}
\label{corners}

In our lattice growth problem, the continuum limiting shape, and even fluctuations of the interface, are just a few key properties. Even if we limit our concern to global characteristics of the interface, we can ask about  
the growth laws for the total number $N_+(t)$ of inner corners and for the total number $N_-(t)$ of outer corners. Using the definition of the growth dynamics we conclude that $\langle N_+\rangle =d V_d/dt$, where $V_d$ is the average volume of the growing crystal. Equation \eqref{Vd_av} then tells us that 
\begin{equation}
\label{N+_growth}
\langle N_+\rangle = dv_d\, t^{d-1}+(d-2+2/z)a_dt^{d-3+2/z}+\ldots
\end{equation}

The dynamics says nothing about the average total number of outer corners. In two dimensions, we can use the obvious topological relation,
\begin{equation}
\label{excess}
N_+(t) - N_-(t) =1 \quad\text{in}~~2D,
\end{equation}
to draw conclusions about the number of outer corners. In two dimensions, we know $v_2=\tfrac{1}{6}$ and $z = \tfrac{3}{2}$, and hence
\begin{equation}
\label{N+_2d}
\langle N_+\rangle = \tfrac{1}{3}t +\tfrac{4}{3} a_2\, t^{1/3}+\ldots
\end{equation}

\begin{figure}[ht!]
\begin{center}
  \includegraphics*[width=0.375\textwidth]{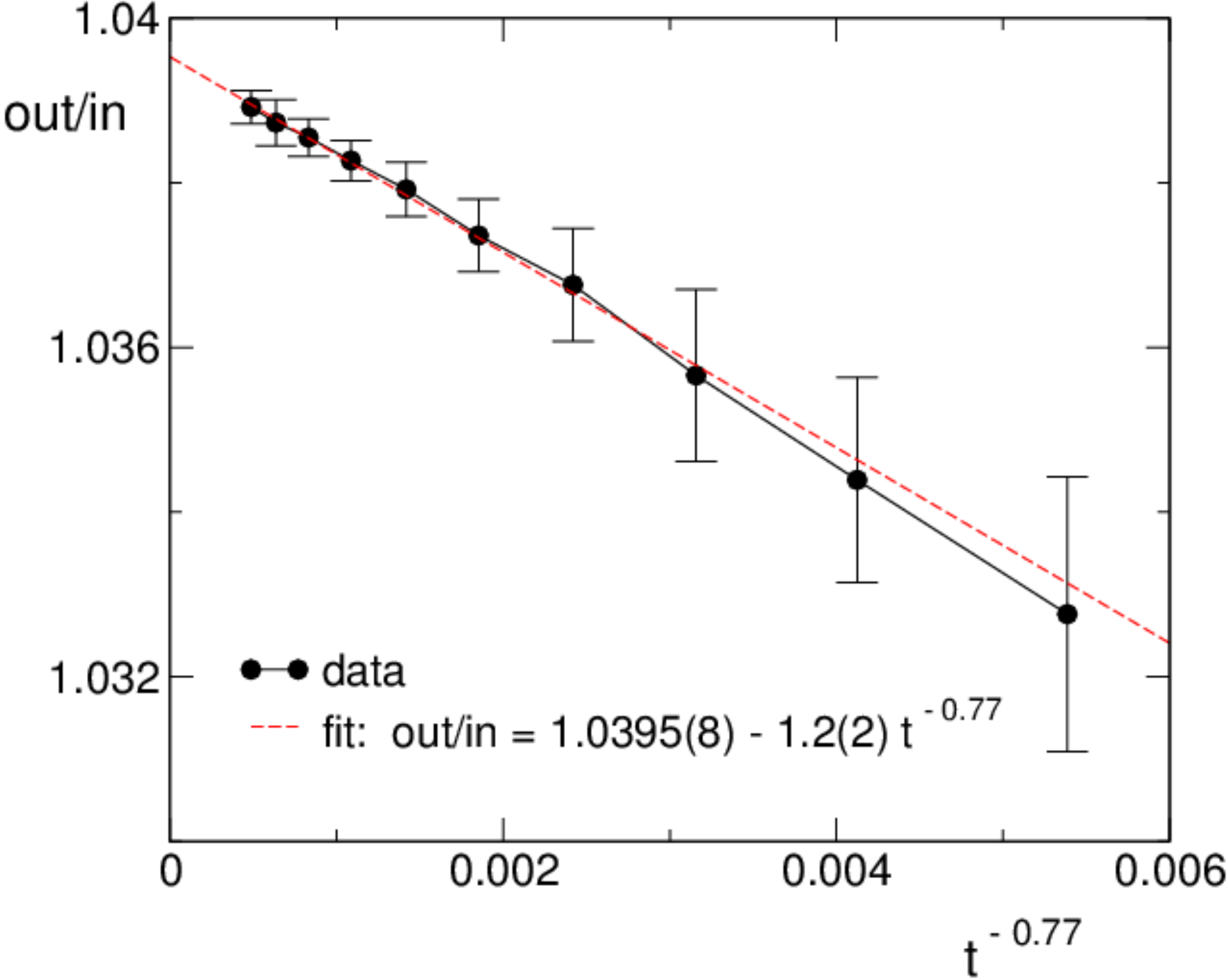}
  \caption{\small The total number of outer corners divided by the total number of inner corners versus $t^{-0.77}$.  
  In the long-time limit, this quantity appears to be greater than 1.}
\label{fig:out_in}
  \end{center}
\end{figure}

In three dimensions, there is no conservation law like \eqref{excess}. For instance, $N_+=6$ and $N_-=3$ for the crystal depicted in Fig.~\ref{Fig:illust_3d}.  We anticipated that generally $N_+>N_-$, and that the average numbers of corners of each kind exhibit the same leading growth.  Intriguingly, our simulation results show that the average total number of outer corners exhibits a faster asymptotic growth (Fig.~\ref{fig:out_in}). Numerically, the leading asymptotic behaviors are
\begin{subequations}
\begin{align}
\label{N+_growth_3d}
\langle N_+\rangle &=  C_+ t^{2}, \qquad C_+  = 0.0442(2)\\
\label{N-_growth_3d}
\langle N_-\rangle &= C_- t^2, \qquad C_-  = 0.0459(2)
\end{align}
\end{subequations}
Our numerical estimate for the amplitude $C_+$ is in good agreement with the theoretical prediction \eqref{N+_growth} according to which $C_+=3v_3=\frac{9\pi^2}{2^{11}}=0.043372285\ldots$. 

\begin{figure}[ht]
\begin{center}
\includegraphics[width=0.375\textwidth]{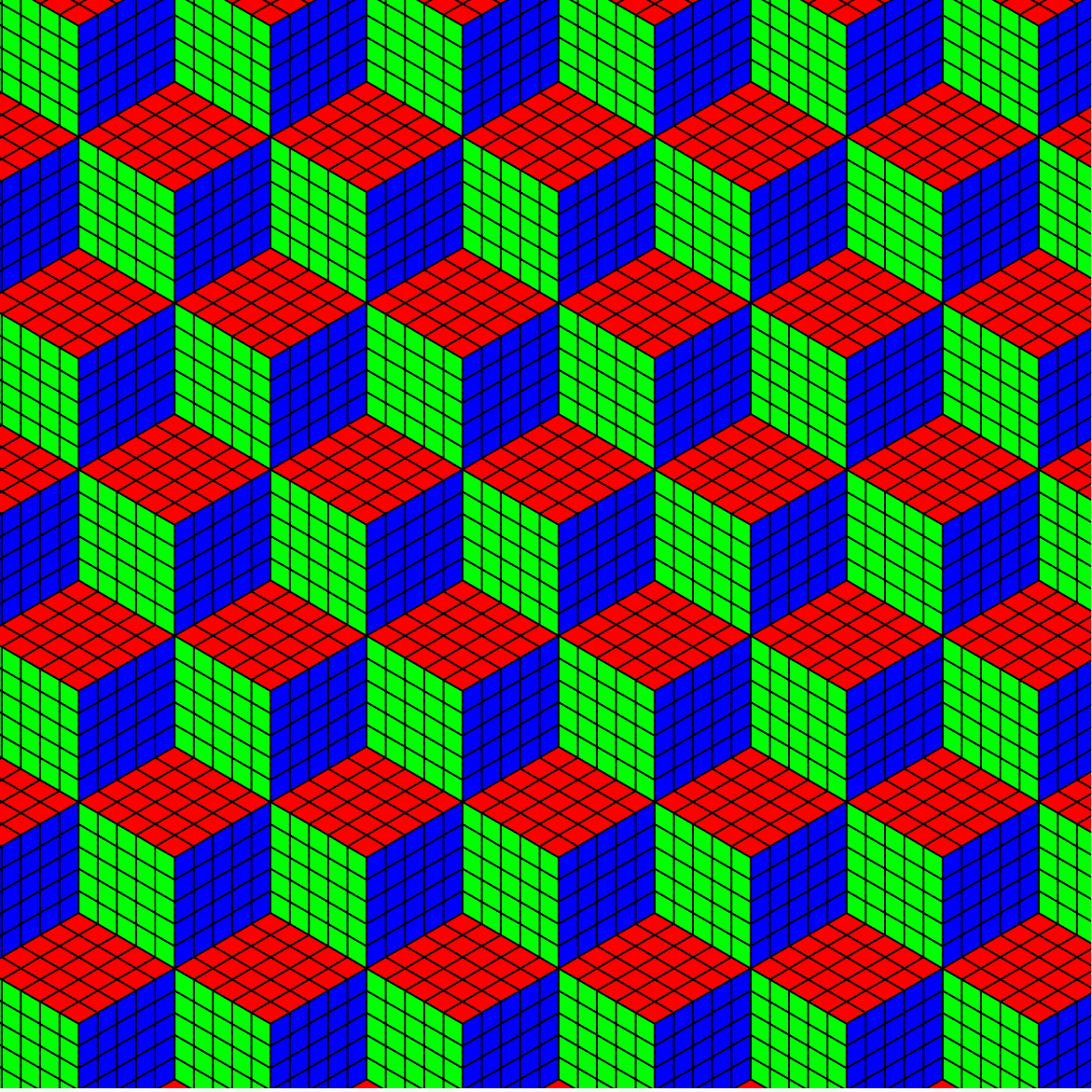}
\vskip 0.1in
\includegraphics[width=0.375\textwidth]{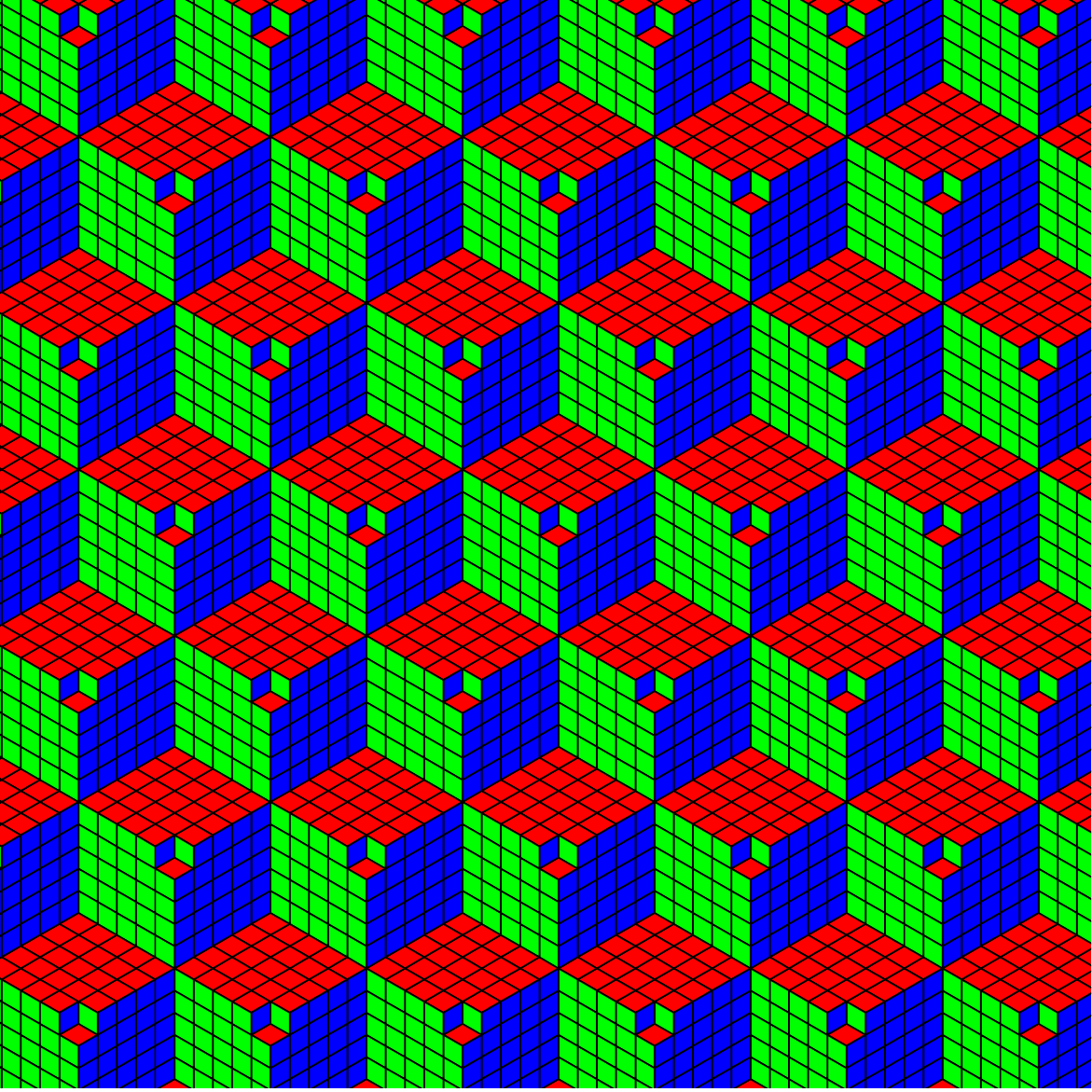}
\caption{\small In the top image, the densities of inner and outer corners on an infinite flat plane are exactly equal.  In the bottom image, we again have an infinite flat plane, but there are now three 
outer corners for every two inner corners.}
\label{Fig:corners}
  \end{center}
\end{figure}

How can the unexpected result $C_->C_+$ be reconciled with the concave curvature of the corner interface (see Fig.~\ref{Fig:shape_3d})?  To aid in understanding, consider Fig.~\ref{Fig:corners}, which shows a flat planar interface, with slopes $z_x=z_y=-1$ in an average sense, that extends to infinity in all directions.  In Fig.~\ref{Fig:corners}~(top), the number of inner corners on the interface exactly equals the number of outer corners.  Since this interface has no average curvature, the equality between inner and outer corners is not so surprising.  Now consider the interface in Fig.~\ref{Fig:corners}~(bottom) after making a small dimple at each outer corner in the top image.  The ratio of outer corners to inner corners is now $3/2$, yet the interface macroscopically still has no curvature!  One sees that the statistics of inner and outer corners cannot be directly inferred from the global shape of an interface.

\section{Discussion}
\label{concl}

We have presented a minimalist model for studying crystal growth in three dimensions.  Elemental cubes are deposited stochastically into an initially empty octant, namely into the inner corners of the growing interface.  At late times when fluctuations become small relative to the typical size of the interface, the interface becomes progressively more deterministic.  We have proposed a hyperbolic partial differential equation \eqref{unique} describing this dominant deterministic limiting shape. This equation has passed through the necessary consistency checks (it has the required symmetry properties, and it reduces to correct evolution equations on the boundaries of the octant). We have solved the evolution equation \eqref{unique} using the Hamilton-Jacobi technique and we have found a very good agreement between analytical results and numerical simulations of the growth process. 

We have also analyzed fluctuations of integral characteristics, e.g. we have expressed the sub-leading correction to the volume of the crystal and the variance of the volume through the KPZ exponents. Even in $1+1$ dimensions, the only case for which these exponents are known, our results are incomplete as we haven't computed the amplitudes. This technically challenging problem may be within the reach of analytical techniques which have been developed in studies of interfaces in $1+1$ dimensions (see e.g. \cite{KPZ_rev,BDJ,J00,Schutz,TW,KM,DG2009a}); perhaps even large deviations could be analytically extracted.   

Overall, we are faced with a dilemma: We haven't derived Eq.~\eqref{unique}, and on the numerical side there is a small (less than $1\%$ in the growth velocity) but persistent discrepancy between the analytical solution of Eq.~\eqref{unique} and simulation results. One possible explanation is that for the $2+1$ dimensional KPZ growth the convergence can be notoriously slow \cite{Wolf,Chate,HH12}. We have shown that one can construct evolution equations passing the consistency checks which provide a better agreement with observed value of the growth velocity. Such equations are ugly, and it seems that if an equation has a right-hand side which is a rational function of the spatial derivatives $z_x$ and $z_y$, the only elegant one which is in a very close proximity with simulations is Eq.~\eqref{unique}. Given the utmost simplicity of the rules of our minimalist growth model, it would be very odd to end up with an ugly equation for the limiting shape. Perhaps the only possible way to overcome the above (admittedly imprecise) arguments is if the true evolution equation has a right-hand side which is a transcendental function of the spatial derivatives $z_x$ and $z_y$ (which reduces to simple rational functions on the boundaries of the octant, i.e. when $z_x=-\infty$ or $z_y=-\infty$). The striking simplicity and accuracy of our conjectural growth equation beautifully match the simplicity of the model under study, which certainly warrants further investigation.

We hope that our analytical results will aid in studies of $2+1$ dimensional KPZ growth.  Relatively little is known about temporal correlations in height fluctuations in these growth models.  Some recent studies on $1+1$ dimensional KPZ growth have revealed slow temporal de-correlation along the characteristic directions of interface growth.  In the polynuclear growth model in $1+1$ dimensions, temporal correlations in height fluctuations have been proven to decay on a time scale $t^{2/3}$ for generic curves in the space-time.  However, along the characteristic directions, temporal correlations decay on a time that scales as $t$, i.e. much more slowly \cite{Ferrari}.  In $2+1$ dimensions, presumably a similar slow de-correlation phenomenon is present, although to our knowledge this has not been tested \cite{slow}.  We now have predictions for the characteristic curves of $3d$ corner growth, and our equations may help facilitate numerical experiments on time-dependent interface fluctuation statistics.

We have offered novel predictions for the sub-leading correction to the average volume \eqref{Vd_av} and for the growth in variance of the volume \eqref{Vd_var} of a growing crystal in three and higher dimensions.  These results should hold generically for all higher-dimensional models that belong to the strong-coupling KPZ universality class.  Indeed, we hope that studies of integral properties of the interface such as the volume or the numbers of inner and outer corners may deliver valuable new insights.

Open problems abound in the field of interface shapes and statistics. A host of problems arises if we reformulate our original problem in terms of the spin-flip dynamics \cite{book}. Consider an Ising ferromagnet with nearest-neighbor interactions endowed with zero-temperature spin-flip dynamics. If we start with a minority phase occupying the octant, the evolution is exactly the same as in our crystal growth process if we additionally postulate that our ferromagnet is in a magnetic field which favors the majority phase. What happens if in addition to the nearest-neighbor interactions we take into account next-nearest-neighbor interactions, or even longer range interactions?  We have recently addressed this problem \cite{KO13} in two dimensions: We have derived evolution equations generalizing Eq.~\eqref{eqn:yt-2d} and found corresponding limiting shapes. There is an infinite series of such equations varying with the range of interactions. Needless to say, nothing is known in three dimensions. 

Even more challenging is to consider the problem without any magnetic field. In our original language, this is equivalent to allowing desorption of elemental cubes from outer corners on the interface.  The desorption proceeds with the same rate as deposition. The interface grows much more slowly than in the pure deposition case, namely the growth is diffusive (that is, the linear size of the interface scales as $\sqrt{t}$). Quantitative results are known only in two dimensions where the exact evolution equation is known and solvable \cite{K12}. In the general case, an evolution equation for unbiased dynamics was proposed in Ref.~\cite{K12}, yet this equation appears analytically intractable. Even numerically this problem has not yet  been studied.  

Finally, we mention equilibrium crystals (also known as Young diagrams) of a fixed volume inside a corner. All crystals with the same volume are equiprobable. A surface of a typical crystal with large volume is close to a limiting shape which has been established both in two \cite{Temp,Vershik} and three \cite{CK_2001,OR_2003} dimensions (see also Refs.~\cite{CLP98,OR07,BFP10} for other three-dimensional versions). In greater than three spatial dimensions, finding the equilibrium limiting shape is a tantalizing mathematical problem. One suspects a connection between equilibrium limiting shapes and the limiting shapes arising in the growth problems (with and without a magnetic field), or perhaps a connection between equations which determine these limiting shapes. One may try to guess equilibrium limiting shapes using the same tricks as before (symmetry constraints and matching to low-dimensional equilibrium limiting shapes in conjunction with a small number of well-chosen numerical clues).

\section*{Acknowledgments}

We are very grateful to K.~Mallick and S.~Redner for numerous discussions and for collaboration at an earlier stage of this work. We are also thankful to A.~Borodin and S.~Ferreira for correspondence, and to I.~Corwin and S.~Prolhac for useful suggestions about fluctuations.  The research of J.O. was supported by NSF Grant No. DMR-1205797.


\begin{thebibliography}{99}
\bibitem{vel} 
      Generally for Eq.~\eqref{multiplicative}, one finds $v=9^c/8^{1+c}$, which is best fit 
      to the simulation data by choosing $c\approx 0.074$; similarly for equations from
      the additive class $8v=1+\frac{c}{9}$.
\end{thebibliography}
\end{document}